\documentclass[twocolumn]{aastex63}

\usepackage{lineno}
\received{2021 April 21}
\revised{2022 January 13}
\accepted{2022 January 20}
\submitjournal{ApJ}

\shortauthors{Blackwell et al.}
\graphicspath{{./}{figures/}}
\usepackage{graphicx}
\usepackage{natbib}

\begin{document}

\defcitealias{Bregman2010}{B10}

\title{The Missing Metal Problem in Galaxy Clusters: Characterizing the Early Enrichment Population}

\correspondingauthor{Anne E. Blackwell}
\email{aeblackw@umich.edu}

\author[0000-0002-0786-7307]{Anne E. Blackwell}
\affiliation{University of Michigan, Ann Arbor, MI 48104, USA}

\author{Joel N. Bregman}
\affiliation{University of Michigan, Ann Arbor, MI 48104, USA}

\author{Steven L. Snowden}
\affiliation{NASA/Goddard Space Flight Center, Lopez Island, WA 98261, USA}

\nocollaboration{3}



\begin{abstract}

Rich and poor galaxy clusters have the same measured halo metallicity, 0.35–0.4 $Z_\odot$, even though they are an order of magnitude apart in stellar fraction, $M_*/M_{gas}$. 
The measured intraclsuter medium (ICM) metallicity in high-mass clusters cannot be explained by the visible stellar population as stars typically make up 3-20\% of the total baryon mass. 
The independence of metallicity of $M_*/M_{gas}$ suggests an external and universal source of metals such as an early enrichment population (EEP). 
Galaxy cluster RX J1416.4+2315, classified as a fossil system, has a stellar fraction of $M_*/M_{gas}=0.054\pm0.018$, and here we improve the halo metallicity determination using archival Chandra and XMM Newton observations. 
We determine the ICM metallicity of RXJ1416 to be $0.303\pm0.053$  $Z_\odot$ within $0.3<R/R_{500}<1$, excluding the central galaxy. 
We combine this measurement with other clusters with a wider range of $M_*/M_{gas}$ resulting in the fit of $Z_{tot}=(0.36\pm0.01)+(0.10\pm0.17)(M_*/M_{gas})$. 
This fit is largely independent of $M_*/M_{gas}$, and shows that for a low $M_*/M_{gas}$ system, the observed stellar population can make only 10-20\% of the total metals. 
We quantify the Fe contribution of the EEP further by adopting a standard Fe yield for visible stellar populations, and find that $Z_{EEP}=(0.36\pm 0.01)–(0.96\pm 0.17)(M_*/M_{gas})$. 
To account for the observed Fe mass, a supernova (SN) rate of $10\pm5$ SNe yr$^{-1}$ (Type Ia) and $40\pm19$ SNe yr$^{-1}$ (core collapse) is required over the redshift range $3<z<10$ for a single galaxy cluster with mass $\sim3\times10^{14}$ $M_\odot$ at z=0.
These SNe might be visible in observations of high-redshift clusters and protoclusters with the James Webb Space Telescope.

\end{abstract}



%%%%%%%%%%%%%%%%%%%%%%%%
%%%    Introduction
%%%%%%%%%%%%%%%%%%%%%%%%
\section{Introduction} \label{sec:intro}
Metals are produced through nucleosynthesis, but the location and time of metal formation by stellar populations throughout cosmic history is still uncertain.
The hot halos of clusters are X-ray emitting extended components in which $>$80 \% of the baryons reside for high-mass clusters ($M_{500} > 10^{14} M_\odot$; \citealp{Andreon2010, Dai2010, KraBor2012}).
These clusters are considered to be closed systems that have retained nearly all their baryons, making their halos the ideal place for studying their stellar and elemental history.

The measured metallicities ($Z$) in the halos of high- and low-mass ($10^{13} M_\odot < M_{500} < 10^{15} M_\odot$) clusters are the same, even though the ratio of stellar mass to gas mass in high-mass clusters (stellar fraction, $M_*/M_{gas}$) is an order of magnitude lower than that of low-mass clusters \citep{Bregman2010, Lui2020}.
\cite{Loew2013} used the standard single-slope Salpeter initial mass function (IMF), applied specifically above the core-collapse mass cutoff, to derive an expected intracluster medium (ICM) metallicity from the current stellar populations of $Z_{Fe,ICM} = 0.106(10f_*/f_{ICM})$ where $f_*/f_{ICM} = M_*/M_{ICM} = M_*/M_{gas}$, using \cite{AndGrev1989} abundances.
For a cluster with $M_*/M_{gas}=0.02$ the expected ICM enrichment from the stellar populations is 0.021 $Z_\odot$.
Considerations such as the efficiency of  Type Ia and core-collapse supernova and a single high-redshift population of stars went into the prediction of observed ICM metallicity, but the measured metallicity is approximately four times greater than what is predicted \citep{Portarini2004, Loew2006, Siv2009, Bulbul2012, Loew2013}.
This is the missing metal conundrum.
%Models show that the metallicity in high $M_*/M_{gas}$ (low mass) clusters is supported by the visible stellar population, but the predicted metallicity of low stellar fraction (high mass) clusters falls short by a factor of 4-8 (\cite{Portarini2004}; \cite{Loew2006}; \cite{Bulbul2012}; \cite{Loew2013};\textbf{more references}).

We consider several possible explanations for the metal conundrum, but find a fault to each.
The first theory is that stars may exist in the ICM that produced a sufficient amount of metals to supply high-mass clusters with the measured ICM metallicity \citep{Renzini&Andreon2014}.
\cite{Siv2009} calculated the average metal contribution from intracluster stars with a standard IMF to be 25\% of the total ICM metallicity within $R_{500}$.
In order for intracluster stars to supply the missing metals, we would expect to find at least four times the intracluster stellar population (based on the metal contribution by \cite{Siv2009}).
Intracluster light in high-mass clusters represents approximately 20\% of the total cluster light \citep[e.g.,][]{Zibetti2005, Krick2007, Giallongo2014, Furnlell2021}.
A population four times larger, a sufficient size to account for the missing metals, would produce at least 80\% of the total cluster light.

The second theory considers a correlation between the IMF of the star-forming clouds at $z \sim 2$ and the final cluster mass \citep{Renzini&Andreon2014}. 
High-mass clusters require an additional source of metals that are not produced by their visible stellar populations. 
These metals could be formed by star-forming clouds at high redshift before the formation of galaxy clusters. 
However, high-mass clusters require a greater source of metals from star-forming clouds than low-mass systems \citep[][and references therein]{Renzini&Andreon2014}. 
The IMF of the star-forming clouds at high redshift would need to compensate for the missing metals by forming more high-mass stars for high final mass systems, and fewer high-mass stars for low final mass systems.
In order for this to be plausible, the star-forming clouds would have to ``know" the final mass of the future cluster, which is unlikely.

A favorable explanation for the metal conundrum is an early enrichment population (EEP); a population of stars that supplied high-mass clusters with most of their measured ICM metals \citep{Elbaz1995, Loew2001}.
This theory has been proposed through the study of halo metallicity evolution as a function of redshift \citep{Mantz2017, Flores2021} and cluster stellar fraction ($M_*/M_{gas}$; \citealp{Bregman2010}).
The EEP would have existed at high redshift, composed of primarily early Population II stars, which dominated during the reionization period $\sim6 < z < 10$, and favored a bottom-light IMF \citep{Fan2006, Loew2013, Planck2016reionization}.
The EEP would have occurred in nascent galaxies and not be prominent as $L^*$ galaxies, or be visible today \citep{WerMer2020}.
This solution avoids problems such as excess intracluster light and halo mass-IMF correlation in high-redshift star-forming clouds.

A few constraints are imposed on the EEP based on current observations. 
The metallicity of high $M_*/M_{gas}$ systems cannot exceed observations, enforcing a maximum EEP metallicity ($Z_{EEP}$).
Limits of the mass of the EEP population ($M_{EEP}$) are imposed as the addition of the EEP and its remnants cannot exceed the cosmic baryon mass.

The need for this population is shown by plotting the $Z_{tot}$ against $M_*/M_{gas}$ \citep[e.g.,][]{Bregman2010}.
We use this methodology to place further constraints on the EEP, such as the supernova (SN) rate and total population mass. 
The high-mass galaxy cluster RX J1416.4+2315 (hereafter RXJ1416) at $z \sim 0.137$ has $M_*/M_{gas} = 0.054 \pm 0.018$ \citep{Khosr2006, Bregman2010, Harrison2012}.
RXJ1416 is also classified as a fossil system (FS), which is an extreme subset of galaxy clusters \citep{Khosr2006}.
FS are suspected to be highly evolved galaxy groups and clusters dominated by a brightest central galaxy (BCG).
The official definition of these systems is generally accepted to be several distinguishing characteristics: (i) a magnitude gap of at least two between the brightest and second brightest members, (ii) both members are within half the virial radius, and (iii) the presence of a diffuse X-ray halo with $L_X \geq 10^{41} h_{70}^{-2}$ erg s$^{-1}$ \citep{Jones2003, Aguerri2011, Aguerri2021}.
Although the definition has been largely agreed upon, many defining factors of FS are unknown or debated, including the formation pathway, the average ICM metallicity, and the radial metallicity trend \citep{Donghia2005, VonBenda2008, Dariush2010}.

Current studies of FS suggest an overabundance of the ICM metallicity \citep[i.e.,][]{Khosr2006}.
This trend lends to the EEP theory, and an early formation of FS as present-day low surface brightness galaxies in FS may have formed alongside the EEP and then evolved to express the defining characteristics of FS.
We study the ICM metallicity of RXJ1416 and include it in the relation of $M_*/M_{gas}$ versus $Z_{tot}$ from \cite{Bregman2010}, from which we can derive the metal contribution from the EEP ($Z_{EEP}$) to the total metallicity ($Z_{tot}$) of the cluster.
We present derived observational quantities of the SNe rate and the total mass of the population from $Z_{EEP}$.

In Section \ref{sec:data} we discuss our handling of the X-ray data from which values presented in Section \ref{sec:spectral_analysis} are extracted.
These values are used to derive observational quantities of the EEP in Section \ref{sec:EEP_quantities}, and we summarize our findings in Section \ref{sec:summary}.

Throughout our study, we adopt a $\Lambda$CDM cosmology with $H_0 = 70$ km s$^{-1}$ Mpc$^{-3}$, a baryon density parameter of $\Omega_b = 0.0455$, dark matter density parameter of $\Omega_{DM} = 0.243$, and the total baryon density parameter is $\Omega_m = 0.288$ \citep{Planck2018cosmo}.

%%%%%%%%%%%%%%%%%%%%%%%%%%%%%%%%%%%%
%%.  Data Reduction and Analysis
%%%%%%%%%%%%%%%%%%%%%%%%%%%%%%%%%%%%
\section{Data Reduction and Analysis}\label{sec:data}

RXJ1416 is a fossil galaxy cluster - a highly relaxed and undisturbed galaxy cluster with an absolute magnitude gap of at least 2 between the brightest and second brightest galaxy \citep{Ponman1994, JoPoFo2000, Jones2003}.
This cluster has been studied and characterized by previous authors \citep{Khosr2006, Khosh2007} using X-ray data and by \cite{Cyp2006} using optical data. 
The first two references provide radial kT measurements for comparison, $M_{200} = 3 (\pm 1) \times 10^{14} M_\odot$ and $R_{500} = 880 kpc$.
We used archival observations of RXJ1416 from XMM-Newton and Chandra to obtain radial kT and $Z_{tot}$ measurements from the system.
Final good-time intervals for data used are presented in Table \ref{tab:gti}.
In addition to XMM-Newton and Chandra data, we included a ROSAT All-Sky Survey (RASS) spectrum to better constrain the diffuse, soft X-ray background.

One further consideration that must be discussed is the value of $M_*$ for RXJ1416.
This value is not reported for the entire cluster, so it must be determined with the data provided.
A reasonable method for determining $M_*$ would be using a provided mass-to-light ratio (M/L) for galaxy clusters.
However, RXJ1416 is an FS, as determined by \cite{Khosr2006}, so typical M/L ratios of galaxy clusters cannot be used as the M/L ratio for FS is known to be higher than that of typical galaxy clusters, but how high is under dispute \cite[e.g.,][]{Vik1999, Sun2004, Khosh2007, Aguerri2021}.
Thus, the $M_*$ of RXJ1416 must be determined for the M/L ratio of the cluster itself.
\cite{Harrison2012} studied the BCG of RXJ1416 and determined $M_{*,BCG} = 1.08 \times 10^{12} M_\odot$ and $L_{BCG} = 3.53 (\pm 0.10) \times 10^{11} L_\odot$.
The M/L ration is then $3.06 (\pm 0.09) M_\odot/L_\odot$ for the BCG.
We use the M/L ration determined and the total \textit{r}-band luminosity within $R_{200}$ for the cluster from \cite{Khosr2006}, $L_{R,R_{200}} = 10^{11.85} L_\odot$, to calculate $M_{*,200} = 2.17(\pm 0.06) \times 10^{12} M_\odot$.
This results in a stellar fraction $M_*/M_{gas} = 0.054 \pm 0.018$.

\begin{table}[]
    \centering
    \caption{Good-time Intervals after cleaning data for XMM-Newton and Chandra detectors.}
    \begin{tabular}{c  c  c  c  c  c}
        XMM Newton & & & & Chandra & \\
        \hline
         OBSID & MOS1 & MOS2 & PN & OBSID & ACIS-I \\
         \hline
         0722140401 & 27 ks & 27 ks & 17.5 ks & 16125 & 31.73 ks \\ 
    \end{tabular}
    \label{tab:gti}
\end{table}

\begin{figure*}
    \centering
    \includegraphics[width=1\linewidth]{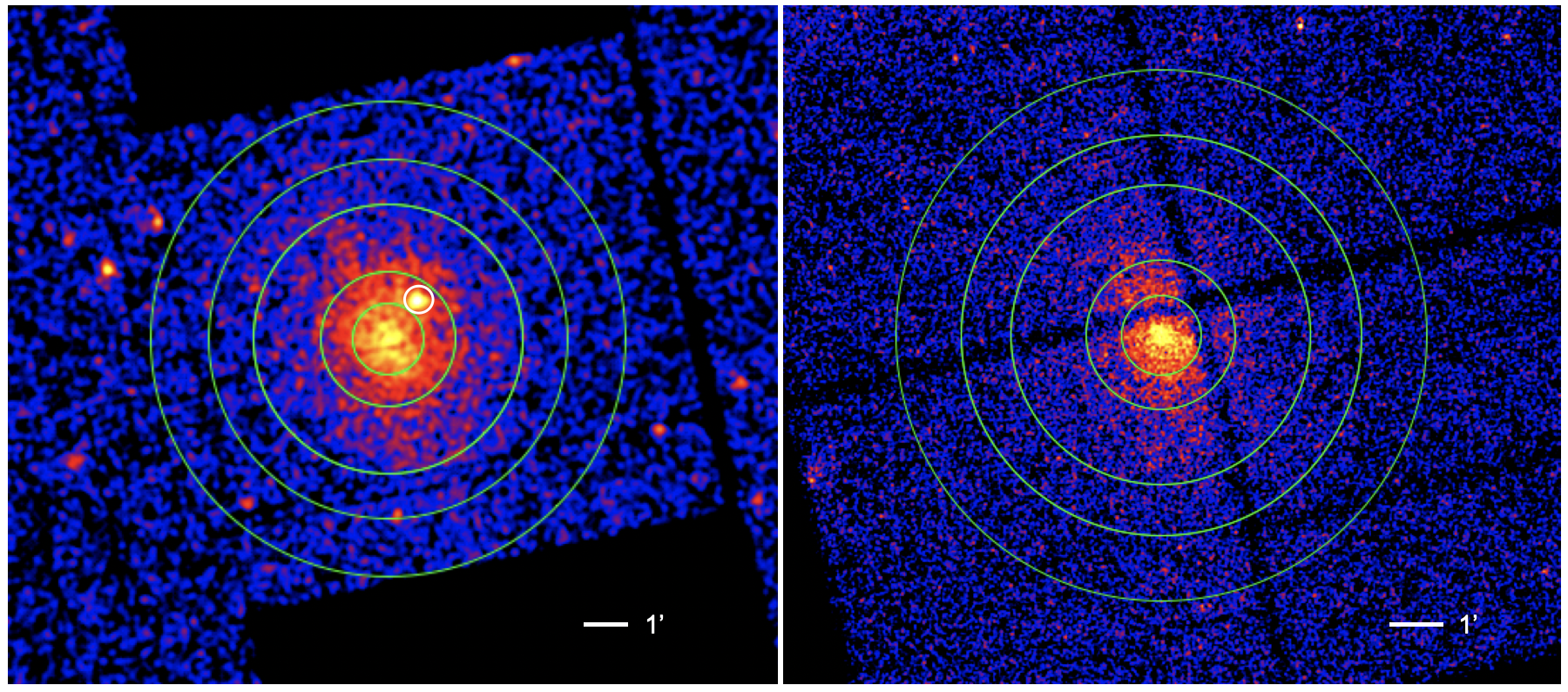}
    \caption{Left: XMM-Newton Mos1 27 ks exposure image of RX J416.4+2315. Not shown, 10$\arcmin$-14$\arcmin$ background region. Right: Chandra ACIS-I chip 30 ks exposure image of RX J1416.4+1315. The same regions are shown on each chip - five annuli from which spectra were extracted and jointly fit. The outermost annuli is of radius $R_{500} = 5.33\arcmin$ \citep{Khosh2007}.}
    \label{fig:rxj_3region}
\end{figure*}

%%%%%%%%%%%%%%%%%%%%%%%%%%%%%%
%%%   XMM Data Reduction
%%%%%%%%%%%%%%%%%%%%%%%%%%%%%%
\subsection{XMM Newton Data Reduction}\label{XMM_Analysis}
RXJ1416 was observed by XMM Newton on four separate occasions. 
Three of the four observations were dominated by soft proton flares, so we limited our study to observation 0722140401 (34 ks exposure before cleaning).
We followed the techniques outlined by \cite{Snowden2008}, the XMM-ESAS Cookbook\footnote{https://heasarc.gsfc.nasa.gov/docs/xmm/esas/cookbook/}, and the Extended Source Analysis Software (xmm-esas, SAS version 18.0.0)\footnote{https://www.cosmos.esa.int/web/xmm-newton/sas} for data extraction and reduction.
We examined the resulting light curves to ensure all flare intervals were excluded from the cleaned observations.
After cleaning, there were 27 ks, 27 ks, and 17.5 ks of good time available for the MOS1, MOS2, and PN instruments, respectively.
We used the filtered data sets for source and background data extraction, and the creation of response matrices and ancillary response files applied to the spectra.

We accounted for two known background and instrumental effects in data reduction and spectral fitting.
The finite point
spread function (PSF) of the XMM-Newton EPIC chips covers multiple annuli in which we extracted spectra of RXJ1416.
This allows for cross talk between the regions; an X-ray photon originating in one extraction annulus on the sky may be detected in another extraction annulus on the chip.
To account for this, we used the SAS command \textit{arfgen} to generate PSF maps between the multiple annuli and account for excess flux generated by X-ray photons detected in annuli from which they do not originate.
The PSF maps were included in the spectral fitting.

A second consideration involves only the PN instrument. 
To improve counting statistics and energy range, we extracted two spectra from the same region.
A low-energy spectrum ($0.3 < E < 2.0$ keV) excluded a low-energy noise tail, and a high-energy spectrum ($1.0 < E < 8.0$ keV) maximized the number of events included at high energies.
During spectral extraction of the PN spectra, we set CCD PATTERN$<= 0$ for the low-energy spectra, and CCD PATTERN$<= 4$ for the high-energy spectra.

XMM Newton EPIC cameras have a field of view of 30$\arcmin$. 
RXJ1416 has $R_{500} = 5.33\arcmin$ and is centered on CCD1 for the MOS instruments, quadrant 1 for the PN.
We defined an on-chip background region annulus of 10$\arcmin$-14$\arcmin$ outside of detectable cluster emission to better constrain the background emission. A prominent background source was noted and excluded during spectral extraction. 
This source is located in defined extraction regions at coordinates (-463,-897) on MOS1 with a radius 2.5$\arcmin$, circled in white in Figure \ref{fig:rxj_3region}.

%%%%%%%%%%%%%%%%%%%%%%%%%%%%%%
%%%  Chandra Data Reduction
%%%%%%%%%%%%%%%%%%%%%%%%%%%%%%
\subsection{Chandra Data Reduction}\label{Chandra Analysis}
RXJ1416 has been observed six times by Chandra. 
Two of the six observations are centered on RXJ1416 and were 81.14 ks (observation ID 16125) and 14.57 ks (observation ID 2024) before filtering.
We used observation 16125 only as the observation was made in VFAINT mode, which has a 5x5 pixel event island instead of a 3x3 event island, allowing for better discrimination between X-rays and cosmic rays.
Additionally, the good time available from 16125, 31.73 ks, is more than double the unfiltered observation time of 2024.
We followed the data reduction procedure outlined by the Extended Source Analysis Guide\footnote{https://cxc.cfa.harvard.edu/ciao/guides/esa.html}. 
All data reduction was performed in CIAO (version 4.12)\footnote{https://cxc.cfa.harvard.edu/ciao/releasenotes/ciao\_4.12\_release.html}, and used the most recent calibration files, CALDB (version 4.9.3)\footnote{http://cxc.harvard.edu/caldb/}.
We reprocessed the raw data to Level 2 event files, from which we extracted spectra and light curves to ensure all flare intervals are properly filtered.
We used the CIAO command \textit{wavedetect} to identify and remove point sources from the image.
The total exposure time available after reprocessing the data was 31.73 ks. 
To properly account for the background, we generated blanksky\footnote{http://cxc.cfa.harvard.edu/ciao/threads/ acisbackground/} background files using an event file excluding point sources and flares.
Background spectra were extracted in the same detector region as the spectra from the blanksky background.

\begin{figure}[h]
    \centering
    \includegraphics[width=0.9\linewidth]{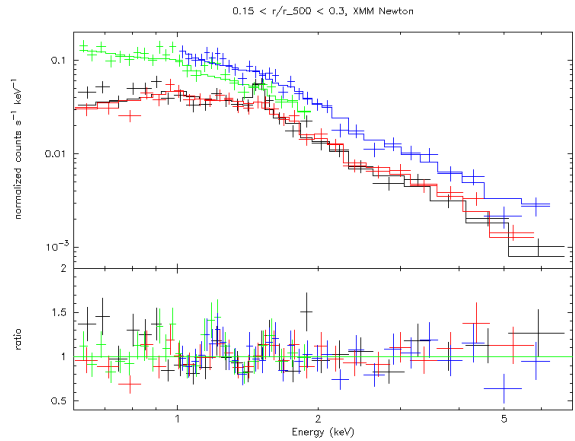}
    \caption{Spectral fit of the $0.15<R/R_{500}<0.30 $ region from XMM Newton. There are four spectra from the region of interest, one each from MOS1 and MOS2, and two from PN, shown in black, red, green (soft energy), and blue (hard energy), respectively.}
    \label{fig:xmm_spec}
\end{figure}

%%%%%%%%%%%%%%%%%%
%%.  Fit Results
%%%%%%%%%%%%%%%%%%
\section{Spectral Analysis}\label{sec:spectral_analysis}
Observational metallicity and plasma temperature trends have been determined and established for numerous galaxy clusters \citep[e.g.,][]{Vik2006, Burns2010, Reiprich2013, WerMer2020}.
Comparison of these trends determined for RXJ1416 against the known trends is necessary to corroborate the determined extraction method and determined values.

For XMM Newton and Chandra, we split up data extraction regions of RXJ1416 for concentric annuli simulated by a beta model to give similar signal-to-noise ratios. 
The defined extraction regions, shown in Figure \ref{fig:rxj_3region}, are $R/R_{500}<0.15$, $0.15<R/R_{500}<0.30$, $0.30<R/R_{500}<0.55$, $0.55<R/R_{500}<0.75$, and $0.75<R/R_{500}<1.0$, where $R_{500} =$880 kpc \citep{Khosh2007}.
We fit the Chandra and XMM Newton spectra independently in the energy range $0.6 - 7.0$ keV using XSPEC (version 12.11.1)\footnote{https://heasarc.gsfc.nasa.gov/xanadu/xspec/}.
Five gaussian lines with fixed energy values are included to account for XMM Newton instrumental lines and solar wind charge exchange.
Normalization and width are initially fit and then frozen. 
A total of three APEC models are used to account for various background X-ray sources, including the Local Hot Bubble, Milky Way halo emission, and the superposition of unresolved background emission.
An additional APEC model is used to constrain the plasma temperature and metallicity of the cluster halo.
Galactic absorption was accounted for using the XSPEC tbabs model set to a fixed column absorption value of $2 \times 10^{20}$ cm$^{-2}$ \citep{Khosh2007}.
We also include constants accounting for calibration differences between instruments and solid angle normalization to allow simultaneous parameter fits using multiple spectral extraction regions. 
A power law is used for the emission from extragalactic active galactic nuclei with index 1.46 and a broken power law for any residual soft proton background emission.
We adopt the solar ratios of \cite{AndGrev1989} for all spectral fits.

\begin{figure}[h]
    \centering
    \includegraphics[width=0.9\linewidth]{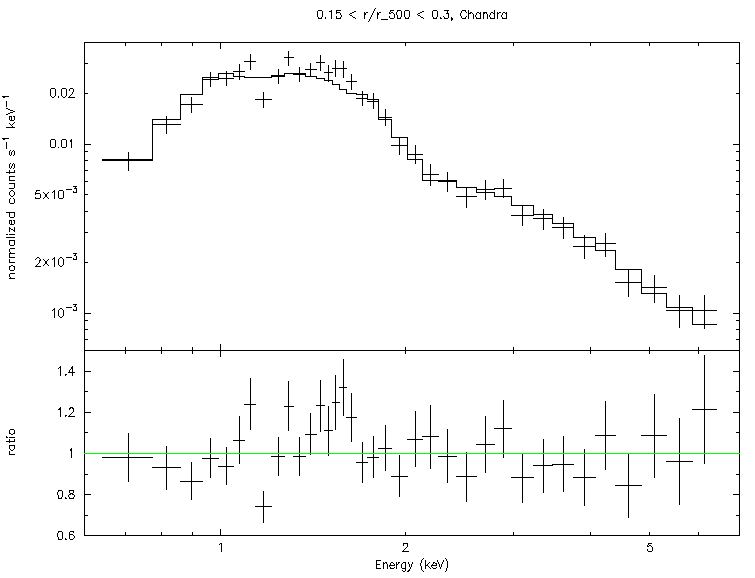}
    \caption{Spectral fit of the $0.15<R/R_{500}<0.30 $ region from Chandra.
    \label{fig:chandra_spec}}
\end{figure}

Fitting a region from XMM Newton includes more spectra than Chandra spectra due to special considerations of the PSF and PN counting statistics.
Each extraction region includes four spectra: MOS1, MOS2, soft-PN, and hard-PN.
Neighboring regions, the on-chip background (10-14'), and a ROSAT spectrum to better constrain the soft thermal emission component are fit simultaneously with the primary region.
The inner ($<0.15$ $R_{500}$) and outermost ($0.75<R/R_{500}<1.0$) extraction regions have a single neighboring region, meaning that 13 spectra are simultaneously fit for each region. 
All other regions have two neighboring extraction regions, so 17 spectra are fit simultaneously.
Figure \ref{fig:xmm_spec} shows an XMM Newton spectral fit of the $0.15 < R/R_{500} < 0.3$ region.
Not shown in this fit are the spectra from the neighboring regions, on-chip background spectra, the on-chip background region, and ROSAT spectra.

We do not have to implement as many considerations when fitting Chandra spectra.
The PSF of the Chandra ACIS-I chip is approximately 1$\arcsec$ \footnote{https://cxc.harvard.edu/proposer/POG/html/chap6.html}, a much smaller area than the width of our defined extraction regions.
A single spectrum is extracted for each defined extraction region and fit simultaneously with a ROSAT spectrum.
Figure \ref{fig:chandra_spec} shows a Chandra spectral fit excluding the ROSAT and background spectra.

%%%%%%%%%%%%%%%%%%%%%%%%%%%%%%
%%%      Metallicity
%%%%%%%%%%%%%%%%%%%%%%%%%%%%%%
\subsection{Metallicity}\label{sec:metallicity}

Metallicity in a relaxed cluster is centrally peaked due to emission from the BCG, then sharply declines, and then remains constant beyond $\sim 0.3$ $R_{500}$ \citep{DeGrande2004, Merniew2018, Lov2019}.
We obtain $Z_{tot}$ measurements in predefined radial bins out to $R_{500}$ for XMM Newton and $0.55$ $R_{500}$ for Chandra.
Spectral fits are not obtained at larger radii for Chandra as high backgrounds dominate the spectra and the signal-to-noise ratios decrease to $< 0.5$ for $Z$ and kT.

We obtain $Z_{tot}$ measurements from extraction regions $0.3<R/R_{500}<0.55$,  $0.55<R/R_{500}<0.75$, and $0.75<R/R_{500}<1$. 
For XMM Newton, we find $Z_{tot}=0.32 \pm 0.073$ $Z_\odot$, $Z_{tot}=0.40\pm0.13$ $Z_\odot$, and $Z_{tot} = 0.14 \pm 0.13$ for the three regions, respectively.
Metallicity measurements are determined only in the region $0.3< R/R_{500} < 0.55$ for Chandra, $Z_{tot} = 0.32 \pm 0.15$ $Z_\odot$.
We calculate a mean of the values weighted by the errors to combine all measurements from Chandra and XMM Newton resulting in a metallicity in annulus $0.3<R/R_{500}<1$ of $0.303 \pm 0.053$ $Z_\odot$.
We note that \cite{Khosh2007} measured an average $Z_{tot} = 0.23 \pm 0.07$ $Z_\odot$ within 0.45$R_{500}$ RXJ1416 obtained by spectral fitting of Chandra data.
This value is within $\sim$1$\sigma$ of the metallicity determined in radial region $0.3-0.55$ $R_{500}$, $0.31 \pm 0.073$.
Because the calibration changes were most likely not included in the errors, we find these measurements in agreement with each other.
All fit results are presented in Table \ref{tab:all_measurements}.

\begin{table*}[]
    \hspace{-1.5cm}
    \centering
    \caption{Metallicity and Plasma Temperature Values from spectral fits in defined regions from both XMM and Chandra.     \textbf{Note.} All errors on spectral quantities provided are 1$\sigma$.}
    \begin{tabular}{c c c c c c}
    \hline
    \textit{XMM Newton} & 0-0.15$R_{500}$ & 0.15 - 0.3 $R_{500}$ & 0.3 - 0.55$R_{500}$ & 0.55 - 0.75$R_{500}$  & 0.75 - 1.0$R_{500}$\\
    \hline
        $Z_{tot}$ ($Z_\odot$)& 0.47 $\pm$ 0.088 & 0.24 $\pm$ 0.079 & 0.32 $\pm$ 0.073 & 0.40 $\pm$ 0.13 & 0.14 $\pm$ 0.13 \\
        $kT$ & 3.46 $\pm$ 0.16 & 3.71 $\pm$ 0.18 & 3.18 $\pm$ 0.15 & 3.01 $\pm$ 0.38 & 2.62 $\pm$ 0.41 \\
        dof & 524 & 588 & 399 & 368 & 418\\
        $chi^2$ & 817 & 903 & 539 & 515 & 547 \\
       \hline
    \textit{Chandra} \\
    \hline
        $Z_{tot}$ ($Z_\odot$)& 0.41 $\pm$ 0.10 & 0.43 $\pm$ 0.06 & 0.32 $\pm$ 0.15 & - & - \\
        $kT$ & 3.30 $\pm$ 0.16 & 4.05 $\pm$ 0.25 & 3.0 $\pm$ 0.82 & - & - \\
        dof & 63 & 33 & 76 & - & - \\
        $chi^2$ & 81 & 49 & 83 & - & - \\
    \hline
    \textit{Weighted Values} \\
    \hline
        $Z_{tot}$ ($Z_\odot$)& 0.44 $\pm$ 0.066 & 0.36 $\pm$ 0.047 & 0.32 $\pm$ 0.065 & - & - \\
        $kT$ & 3.47 $\pm$ 0.10 & 3.82 $\pm$ 0.14 & 3.17 $\pm$ 0.14 & - & - \\
    \end{tabular}
    \label{tab:all_measurements}
\end{table*}

%%%%%%%%%%%%%%%%%%%%%%%%%%%%%%
%%%     kT Radial Trend
%%%%%%%%%%%%%%%%%%%%%%%%%%%%%%
\subsection{kT Radial Trend}

\begin{figure}
    \centering
    \includegraphics[width=0.9\linewidth]{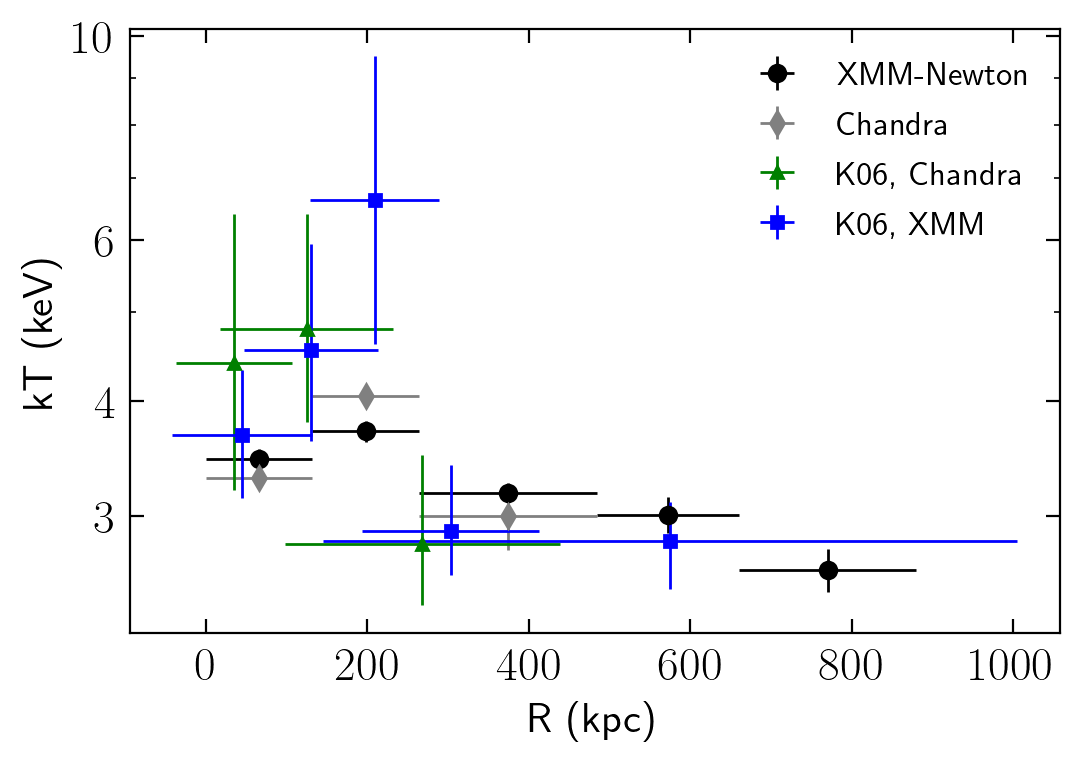}
    \caption{Comparison of radial kT measurements. We find agreement within 2$\sigma$ of all measurements, which may be accounted for by improvements in calibration since 2006.  Here we reference the \cite{Khosr2006} published values as K06.
    \label{fig:kT_all}}
\end{figure}

\cite{Khosr2006} studied RXJ1416 using XMM Newton and Chandra observations.
The authors determined a radial deprojected temperature profile for RXJ1416.
We find agreement of the kT values within a 90\% confidence interval for all values with those presented in \cite{Khosr2006}, as shown in Figure \ref{fig:kT_all}.

\begin{figure}
    \centering
    \includegraphics[width=0.9\linewidth]{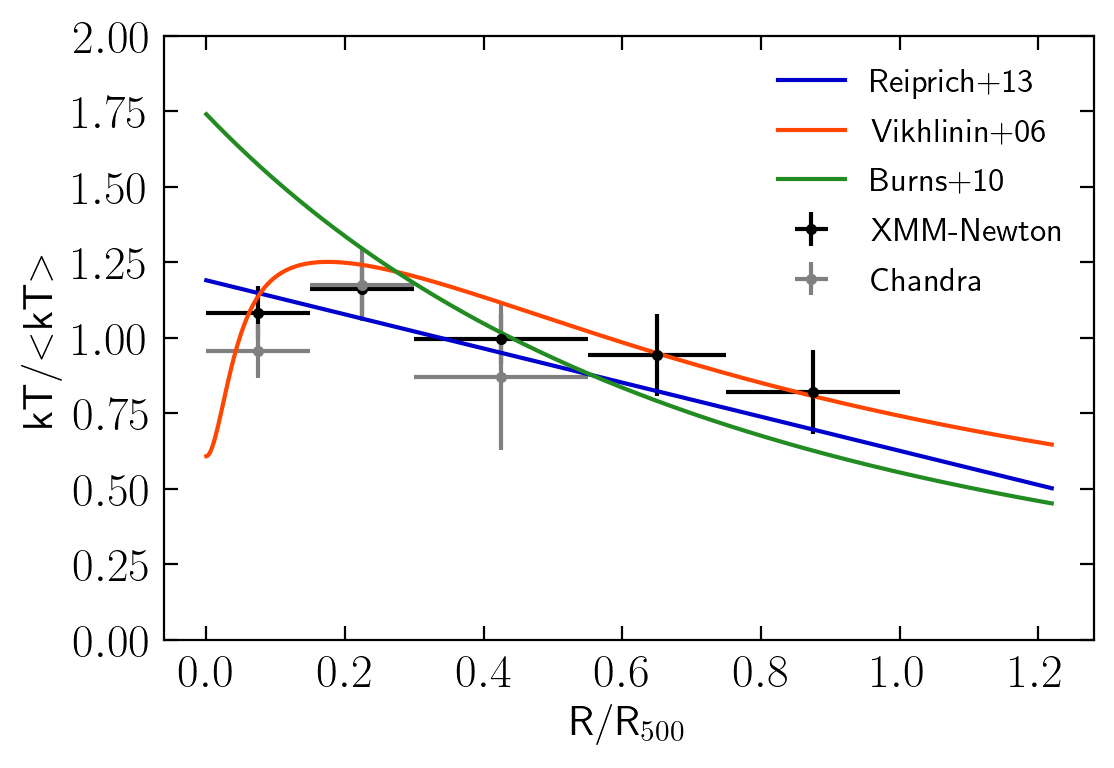}
    \caption{Plasma temperature radial trend of RXJ1416. We compare our XMM Newton and Chandra measurements to published trends \citep{Vik2006, Burns2010, Reiprich2013}.
    \label{fig:kT_profile}}
\end{figure}

To further corroborate our radial kT trend, we compare determined values with published kT trends \citep{Khosr2006, Vik2006, Burns2010, Reiprich2013}.
Figure \ref{fig:kT_profile} shows the resulting radial trends.
We find agreement with the projected temperature model from Equation 9 in \cite{Vik2006}.

%%%%%%%%%%%%%%%%%%%%%%%%%%%%%%%%%%%%%
%%.  Characterizing the EEP from RXJ
%%%%%%%%%%%%%%%%%%%%%%%%%%%%%%%%%%%%%
\section{Characterizing the EEP}\label{sec:EEP_quantities}
The observed metal abundance in the halos of rich galaxy clusters cannot be explained by the visible stellar populations.
We adopt the EEP theory to explain the measured over-abundance of metals. 
The EEP is theorized to be composed of primarily Population II stars that dominate during the reionization period $\sim 6<z<10$, and supplying the metals to enrich subsequent galaxy clusters.

Domination of the Population II generation supplies a high-z limit to the existence of the EEP, but the lower limit requires further consideration. 
The gas must have been enriched before the formation of galaxy clusters ($z \gtrsim 2$) as to supply them with metals.
Simulations by \citet{Biffi2018a} and \citet{Biffi2018b}, show that a flat radial metal distribution in a galaxy indicates that metals escaped their host galaxies around $2 < z < 3$.
We consider a redshift range of $3 < z < 10$ for the existence and calculations of the EEP characteristics.

\cite{Bregman2010} (hereafter \citetalias{Bregman2010}) studied the impact of the EEP based on the relation of $Z_{tot}$ and $M_*/M_{gas}$.
If we assume $M_*/M_{gas}$ to be the controlling factor of cluster properties such as the abundance of the EEP, we can write an expression allowing us to constrain the EEP contribution from known measurable quantities, as expressed in Equation \ref{eq:Z_tot}.

\begin{equation}
\vspace{0.4cm}
Z_{tot} = Z_* + Z_{EEP} = a_1(\frac{M_*}{M_{gas}}) + a_2(\frac{M_{EEP}}{M_{gas}})
    \label{eq:Z_tot}
\end{equation}

\begin{figure}
    \centering
    \includegraphics[width=0.9\linewidth]{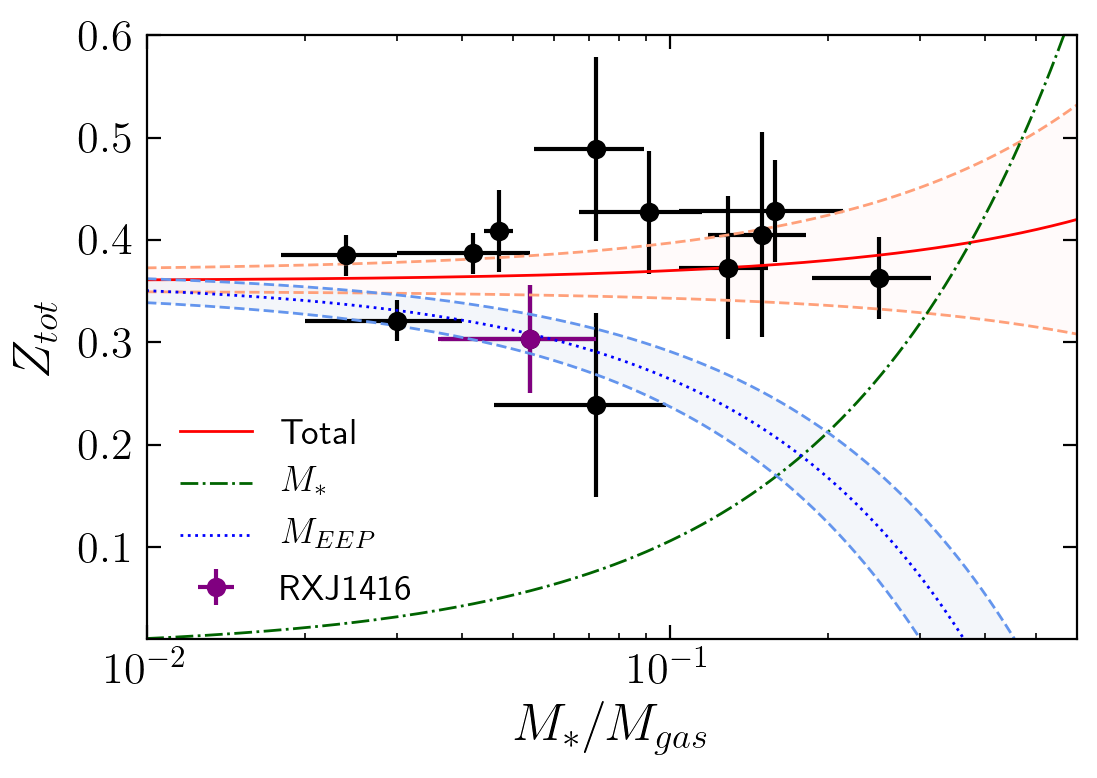}
    \caption{Fit components for \citetalias{Bregman2010} data with RXJ1416 data point. The red line shows the fit between $Z_{tot}$ and $M_*/M_{gas}$ with 1$\sigma$ errors, fit $Z_{tot}=(0.36 \pm 0.01)+(0.10 \pm 0.17)(M_*/M_{gas})$, and the green line is the contribution from $Z_*$. The EEP contribution, shown by the blue line, is the difference between the two.}
    \label{fig:fin_fit}
\end{figure}

Factors $a_1$ and $a_2$ are the metal yield from the visible stars with mass $M_*$, and EEP with mass $M_{EEP}$, respectively.
The contribution of the EEP decreases with increasing $M_*/M_{gas}$.
We use a Markov chain Monte Carlo method to fit this relation to the \citetalias{Bregman2010} data, resulting in a final relationship of $Z_{tot}=(0.37\pm0.01)+(0.13\pm0.17)(M_*/M_{gas})$.
Including the data point for RXJ1416 in the fit from \citetalias{Bregman2010} (Figure \ref{fig:fin_fit}) results in no significant changes to the overall relationship: $Z_{tot}=(0.36\pm0.01)+(0.10 \pm 0.17)(M_*/M_{gas})$.

%Galaxy cluster RXJ1416 has a low stellar fraction of $0.027 \pm 0.009$ \citep{Khosr2006, Bregman2010, Lagana2013}.
%This data point will have leverage on the overall slope and intercept of the line relating $Z_{tot}$ and $M_*/M_{gas}$.
%Figure \ref{fig:fin_fit} shows the total fit to the \citetalias{Bregman2010} data with RXJ1416 ($Z_{tot} = 0.303 \pm 0.053$ $Z_\odot$), resulting in the fit $Z_{tot} = (0.36 \pm 0.01) + (0.12^{+0.17}_{-0.16})(M_*/M_{gas})$.
%The inclusion of RXJ1416 results in no significant changes to the \citetalias{Bregman2010} fit.

The total metallicity in the cluster is the sum of $Z_*$ and $Z_{EEP}$.
\cite{Loew2013} derived a relation for the expected ICM enrichment (specifically, the Fe mass returned) from the evolved stellar populations as a function of $f_*/f_{ICM} = M_*/M_{gas}$ as $Z_* = 1.06 (M_*/M_{gas})$.
We adopt this relation for $Z_*$ versus $M_*/M_{gas}$.
The remaining contribution to the metallicity by the EEP is found by subtracting the fit for $Z_*$ from the total fit for $Z_{tot}$ versus $M_*/M_{gas}$: $Z_{EEP}=(0.36\pm0.01)-(0.96\pm0.17)(M_*/M_{gas})$.
We move forward with constraining the EEP population based on this fit.

%%%%%%%%%%%%%%%%%%%%%%%%%%%%%%%%%%%%%
%%.          EEP SNeR
%%%%%%%%%%%%%%%%%%%%%%%%%%%%%%%%%%%%%
\subsection{EEP Supernova Rate}\label{sec:sne}

We begin our calculation of the EEP expected SN rate by adopting the metallicity contribution and mass for RXJ1416.
From Figure \ref{fig:fin_fit} we determine the EEP metallicity contribution to be $Z_{EEP} = 0.245 \pm 0.056$ $Z_\odot$ for RXJ1416,

\begin{equation}
\label{eq:mass}
    M(r < {R_\Delta}_c) = \frac{4}{3} \pi R_\Delta^3 \Delta \rho_c
\end{equation}

We use Equation \ref{eq:mass} to determine a gravitational mass within $R_{500}$ of $2.9 (\pm 1.0) \times 10^{14} M_\odot$ from the published gravitational mass of RXJ1416 within $R_{200}$, $3.1 (\pm 1.0) \times 10^{14} M_\odot$ \citep{Cyp2006, Khosr2006}.
Here $R_\Delta$ is the radius of the overdensity, $\Delta$ is the overdensity constant, and $\rho_c$ is the critical density of the universe at the redshift of the object.
We then multiply the gravitating mass within $R_{500}$ by the gas fraction for RXJ1416 (0.13$\pm$0.06) to obtain $M_{g,500} = 3.75(\pm 1.3) \times 10^{13}$ $M_\odot$.
The baryonic mass for RXJ1416 is the addition of the stellar and gas mass, resulting in $M_b = 3.96(\pm1.3)\times 10^{13}$ $M_\odot$. 
The baryonic mass multiplied by the fractional value of Fe in the Sun (0.0019) and $Z_{EEP}$ ($0.245 \pm 0.056$ $Z_\odot$) at $M_*/M_{gas} = 0.054$ results in the Fe mass held in baryons, $M_{Fe} = 1.85 (\pm 0.85) \times 10^{10}$ $M_\odot$ \citep{AndGrev1989}.

We consider three types of SN for the creation of Fe in the cluster: SNe Ia, core collapse (CC), and pair-instability (PISN).
SNe Ia and CC produce 0.743 $M_\odot$, and 0.0825 $M_\odot$ on average, respectively, for a standard Salpeter IMF \citep{TNY1986, TNH1996, Kobay2006, Loew2013}.
PISN occur in the most massive stars.
The minimum cutoff of PISN is uncertain, with \cite{Chat2021} reporting a minimum mass of $M > $65 $M_\odot$, and \cite{Morsony2014} and \cite{Heger2002} adopting $140 M_\odot$.
We will adopt a minimum mass of 96 $M_\odot$ as \cite{Woosley2017} found that a PISN of initial mass 96 $M_\odot$ produces 0.045 $M_\odot$ of Fe.
In our assumed IMF from \cite{Loew2013}, 3.4\% of the total population will exist within the high-mass range 95-150 $M_\odot$.
If all of the stars with mass 95-150 $M_\odot$ result in PISN, the relative metal contribution will be 0.6\% of the total Fe production, considering mass intervals of 1-8$M_\odot$ for SNe Ia, 8-95 $M_\odot$ for CC, and 95-150$M_\odot$ for PISN.
The contribution of PISN to the total Fe yield for the IMF considered is insignificant relative to the Ia SNe contribution, thus we limit our continued study to just SNe Ia and CC.
We refer to \cite{Morsony2014} for an expanded discussion of the different IMFs required to produce the observed ICM metal enrichment.

The total metal contribution from Ia and CC can be determined by considering the relative contributions from each type of SNe. 
We adopt the same parameters as \cite{Loew2013} for a diet-Salpeter IMF where the relative fraction of Ia SNe is $f_{Ia} \sim 20 \%$.
Assuming the remaining SNe are CC ($\sim 80 \%$), the average Fe contribution from an SNe is $0.215 M_\odot$.
The total number of SNe required to produce $1.85 (\pm 0.85) \times 10^{10} M_\odot$ of Fe is then $8.6 (\pm 3.9) \times 10^{10}$ SNe.

The redshift over which we expect the EEP to have existed is not well constrained, $\sim 3 < z < 10$, equivalent to $1.69$ Gyr.
The expected number of SNe (Ia + CC) in a single galaxy cluster is $50.9 (\pm 23.5)$ SNe yr$^{-1}$.
This can be broken down into the expected total SNe Ia, $10.2 \pm 4.7$ SNe yr$^{-1}$, and CC $40.8 \pm 18.8$ SNe yr$^{-1}$ for a single galaxy cluster over $3 < z < 10$.

\begin{figure}
    \centering
    \includegraphics[width=0.9\linewidth]{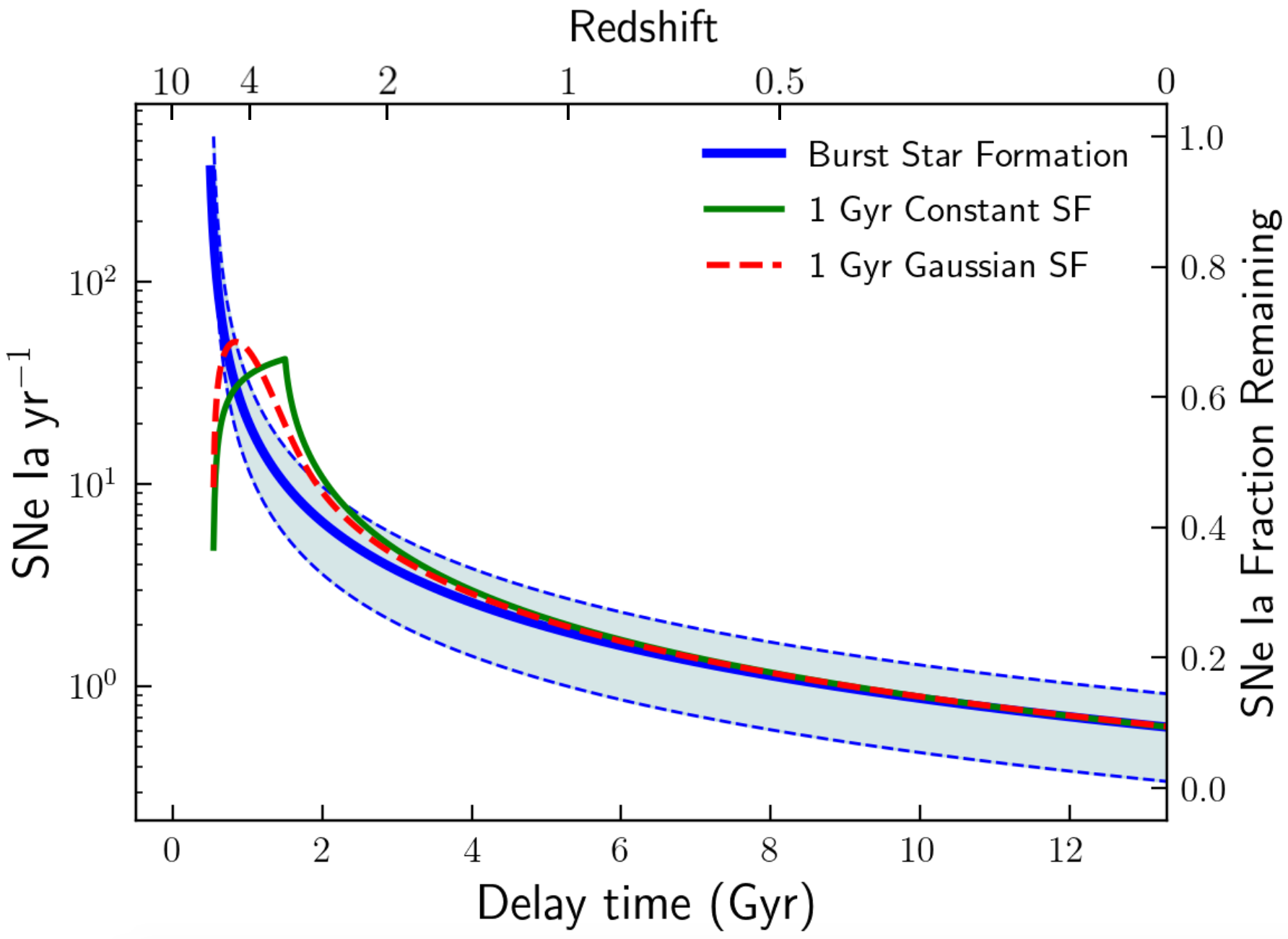}
    \caption{Expected SNe Ia rate modeled by a DTD of $t^{-1.1}$  \citep{Maoz2012}. The blue line shows the expected rate for RXJ1416 since the initial formation of the population at time$=0$. The green and dashed red lines show the effect on the observed SNe Ia rate for a period of constant star formation during z=6-10 (green) and for a Gaussian star formation period during z=6-10.}
    \label{fig:sne_rate}
\end{figure}

Supernovae Ia associated with the EEP may be observable in galaxy clusters today due to this time delay.
The delay-time distribution (DTD) of SNe Ia is best fit by $\sim$t$^{-1}$ (Equation 13 in \cite{Maoz2012}).
We determine a normalization factor to this trend by integrating over time and setting the equation equal to the total number of SNe Ia predicted for RXJ1416.
Figure \ref{fig:sne_rate} shows the calculated SNe Ia DTD for RXJ1416 compared to Equation 13 from \cite{Maoz2012}, determined by observations.
Our predicted SNe Ia trend lies below the trend determined from observations.
It is possible that the high-redshift SNe Ia already observed originated from the EEP, but were not identified.

The trend of SNe Ia as a function of redshift may also provide insight into the star formation rate of the population.
Figure \ref{fig:sne_rate} also demonstrates the differing formation rates of the EEP convolved with the SNe Ia DTD. 
A burst of star formation will result in a simple decay proportional to t$^{-1}$.
Any extended periods of star formation such as a constant rate and gaussian distribution will widen the initial part of the power-law DTD.
The next generation of telescopes, such as the James Webb Space Telescope (JWST), is needed to probe deep into space and monitor the SNe Ia rate of protoclusters.
Efforts to determine the observation limit of SNe Ia with the JWST have already been undertaken, with estimates extending to z=4 \citep{Reg2019}.

\subsection{EEP Yield}\label{sec:a2}
The total halo metallicity $Z_{tot}$ is the sum of the metals produced by the visible stellar populations, $Z_*$ and the EEP, $Z_{EEP}$, Equation \ref{eq:Z_tot}.
Each metallicity contribution can be broken down further into a yield factor, $a$, multiplied by the stellar fraction of the respective population. 
For the visible stellar population, this factor is $Z_* = a_1 (M_*/M_{gas})$.
\cite{Loew2013} developed this relation for clusters as $Z_* = 1.06 (M_*/M_{gas})$.

The metallicity contribution from the EEP can be written as $Z_{EEP} = a_2(M_{EEP}/M_{gas})$ where $a_2$ is the metal yield, and $M_{EEP}/M_{gas}$ is the EEP stellar fraction.
The value $a_2$ can be determined by first solving for $M_{EEP}$, the total mass of EEP stars, and using Equation \ref{eq:Z_tot} to solve for $a_2$.

The total mass of the EEP needed to account for $Z_{EEP}$ may be determined by dividing the total number of EEP SNe (CC+Ia) by the number of SNe per star formed ($\eta^{SN}$) (Equation \ref{eq:meep}). 
We adopt the same diet-Salpeter IMF considered in \cite{Loew2013} which  derives an $\eta^{SN}$ of 0.01.
The total number of EEP SNe calculated in Section \ref{sec:sne} is $8.6 (\pm 3.9) \times 10^{10}$ SNe.
Dividing the total EEP SNe by $\eta^{SN}$ results in $M_{EEP} = 8.6(\pm 3.9) \times 10^{12}$ $M_\odot$.
For the values relevant to RXJ1416, we derive $a_2 = 1.14 \pm 0.16$.
We note that the uncertainties derived for $a_2$ are based on the assumed values of the SNe fraction ($f_{ia}$, and $f_{CC}$), SNe yields ($y_{ia}$, and $y_{CC}$) and number of SNe formed per star formed ($\eta^{SN}$).
The uncertainty is thus independent of any measured value. 
The final equation can be written as $Z_{tot} = 1.06 (M_*/M_{gas}) + 1.14 (M_{EEP}/M_{gas})$, from which $M_{EEP}$ may be found for any system with a measured $M_*$, $M_{gas}$, $Z_{tot}$, and assumed $a_1$ and $a_2$.

\begin{equation}
    M_{EEP} = \frac{SNe (CC+Ia)}{\eta^{SN}}
\label{eq:meep}
\end{equation}

%%%%%%%%%%%%%%%%%%
%%.  Future Work
%%%%%%%%%%%%%%%%%%
\subsection{Future Improvements of EEP Constraints}\label{sec:improvements}

The relation of $Z_{tot}$ vs $M_*/M_{gas}$ was first studied by \citetalias{Bregman2010} who concluded $Z_{tot} = (0.37 \pm 0.01) + (0.13 \pm 0.17)(M_*/M_{gas})$.
Metallicity and stellar fraction values used to determine this relation were obtained from a variety of sources, likely adding to the scatter.
Since \citetalias{Bregman2010}, more data have been made publicly available and calibration files have been improved.
High-precision work may also be applied to the best (brightest and deeply observed) systems, from which one may identify classes of groups or clusters that are statistically different.

We plan to better constrain the EEP contribution and properties by expanding upon the relation between $Z_{tot}$ and $M_*/M_{gas}$.
Data reduction and analysis techniques developed and tested for object RXJ1416 will be implemented for a survey of 27 galaxy groups and clusters.
These selected sources also have $M_*$ and $M_{gas}$ measurements from a single source \citep{Lagana2013}.
Simulations of uniform $Z_{tot}$ measurements of the 27 groups and clusters show a consistent improvement on the slope error by a factor of 2.5-3.

This reduction in errors propagates to reduced errors on the SNe rate and total population mass of the EEP.
Better constraints on this population will allow for a more confident determination of the existence of EEP in future next-generation observations.

%%%%%%%%%%%%%%%%%%
%%.  Conclusion
%%%%%%%%%%%%%%%%%%
\section{Summary}\label{sec:summary}
High-mass galaxy clusters do not have adequate visible stellar populations to produce the measured halo metallicity.
This is the the missing metal conundrum \citep[e.g.,][]{Loew2013}.
A proposed solution to the metal conundrum is a population of high-mass stars that produced most of metals seen in present-day galaxy clusters, the early enrichment population. 
The contribution of the EEP to measured cluster metallicity can be determined by studying the relation between $Z_{tot}$ and $M_*/M_{gas}$ \citepalias{Bregman2010}.
A fraction of the total metal comes from the visible stellar population ($Z_*$), and the remainder comes from the EEP ($Z_{EEP}$).

We improve on the halo metallicity determination of RXJ1416 and break down the total halo metallicity into the contribution from $Z_*$ and $Z_{EEP}$.
Measurement of $Z_{EEP}$ allows for constraints to be imposed on the EEP SNe rate.
The primary calculations are summarized below.
\begin{enumerate}
    \item The RXJ1416 halo metallicity was measured to be $0.303 \pm 0.053$ $Z_\odot$ within $0.3 < R/R_{500} < 0.75$ using archival Chandra and XMM Newton data.
    
    \item Incorporating the halo metallicity of RXJ1416 to the \citetalias{Bregman2010} $Z_{tot}$ versus $M_*/M_{gas}$ relation confirms but does not change the fit significantly: $Z_{tot} = (0.36 \pm 0.01) + (0.10 \pm 0.17) (M_*/M_{gas})$.
    
    \item The EEP metallicity weight, $a_2$, is calculated to be $1.14\pm0.16$. We rewrite the equation for $Z_{tot}$ to incorporate the yields of the two contributing stellar populations: $Z_{tot}=Z_*+Z_{EEP}=1.06(M_*/M_{gas})+1.14(M_{EEP}/M_{gas})$.

    \item  The visible stellar population RXJ1416 halo metallicty contribution ($Z_*$) is $0.057 \pm 0.019$ Z$_\odot$, and the EEP contribution ($Z_{EEP}$) is $0.245 \pm 0.056$ Z$_\odot$.
    
    \item An SNe rate $10.2 \pm 4.7$ Ia SNe yr$^{-1}$ and $40.8 \pm 18.8$ CC SNe yr$^{-1}$ for a single protocluster over the redshift range $3<z<10$ is required to produce $Z_{EEP}$ in RXJ1416.

\end{enumerate}

Future work will improve upon the current fit by extending this analysis to 27 additional galaxy groups and clusters. 
This sample has a range of $0.024 - 0.48$ $M_*/M_{gas}$, all obtained from a single source \citep{Lagana2013}.
The methodology described in Section \ref{sec:data} will be used to determine halo metallicities for each $M_*/M_{gas}$ and $Z_{tot}$.
These new halo metallicities will be used to rederive the contribution and characteristics of the EEP, decreasing the error by a factor of $\sim$3.
In addition to further study of the observational data, simulations of large-scale structures are needed that take the EEP into account.
Decreasing the error in this relation propagates a decreased error to the SNe rate from which we may determine the existence of the EEP in next-generation telescope observations.

%%%%%%%%%%%%%%%%%%
%%.  Conclusion
%%%%%%%%%%%%%%%%%%
\textit{Acknowledgements}. We deeply thank Alexey Vikhlinin for his help on the reduction of the spectroscopic Chandra data and derived value comparisons. 
We would like to thank Cameron Pratt, Jiangtao Li, Zhijie Qu, Rui Huang, and Hang Yang for thoughtful discussion over the course of this research project.
The authors would like to gratefully acknowledge support for this program through the NASA ADAP award AWD 012791.

\bibliography{main}{}
\bibliographystyle{aasjournal}



\end{document}